\documentclass[12pt,draftcls,onecolumn]{IEEEtran}
\usepackage{amsmath}
\usepackage{theorem}
\usepackage{amsmath}
\usepackage{mathrsfs}
\usepackage{accents}
\usepackage{url}
\usepackage{algorithm}
\usepackage{algorithmic}
\usepackage{epsfig}
\usepackage{amsfonts}
\usepackage{amssymb}
\usepackage{mathrsfs}
\usepackage{euscript}
\usepackage{graphicx}
\usepackage{multirow}
\usepackage{cite}
\newtheorem{theorem}{Theorem}

\usepackage[T1]{fontenc}
\usepackage[USenglish,UKenglish,french,spanish,italian]{babel}
\usepackage[nodayofweek,level]{datetime}
\newcommand{\mydate}{\formatdate{7}{4}{2015}}

\graphicspath{{./}{figures/}}

\ifCLASSINFOpdf
\else
\fi
\begin{document}

\begin{titlepage}

\begin{tabular}{l        r}

\includegraphics[bb=20bp 00bp 500bp 450bp,clip,scale=0.3]{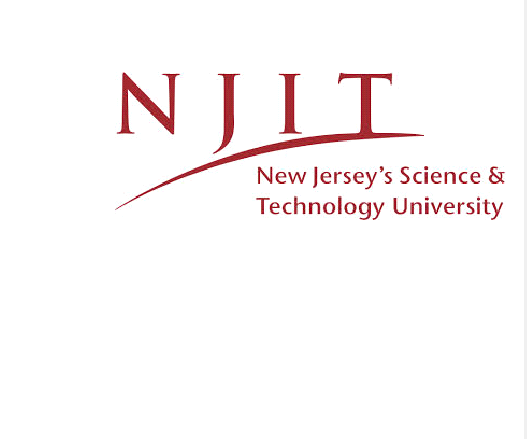} \hspace{6cm} & \includegraphics[bb=0bp -200bp 500bp 550bp,clip,scale=0.2]{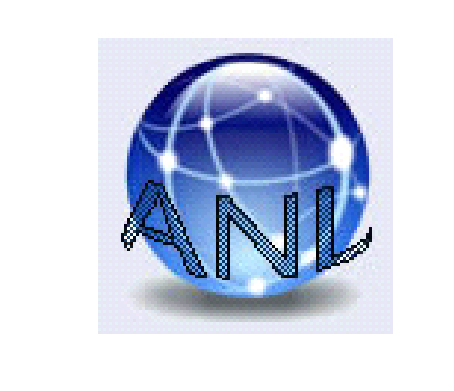}

\end{tabular}

\begin{center}

\textsc{\LARGE Profit Maximization for Geographical Dispersed Green Data Centers}\\[1.5cm]

{\Large \textsc{Abbas Kiani}}\\ 
{\Large \textsc{Nirwan Ansari}}\\ 
[2cm]

{}
{\textsc{TR-ANL-2015-002}\\
\selectlanguage{USenglish}
\large \mydate} \\[3cm]

{\textsc{Advanced Networking Laboratory}}\\
{\textsc{Department of Electrical and Computer Engineering}}\\
{\textsc{New Jersy Institute of Technology}}\\[1.5cm]
\vfill

\end{center}

\end{titlepage}


\selectlanguage{USenglish}
\begin{abstract}
This paper aims at maximizing the profit associated with running geographically dispersed green data centers, which offer multiple classes of service. To this end, we formulate an optimization framework which relies on the accuracy of the G/D/1 queue in characterizing the workload distribution, and taps on the
merits of the workload decomposition into green and brown workload served by green and brown energy resources.
Moreover, we take into account of not only the Service Level Agreements (SLAs) between the data centers and clients but also different deregulated electricity markets of data centers located at different regions. We prove the convexity of our optimization problem and the performance of the proposed workload distribution strategy is evaluated via simulations.
\end{abstract}


\section{Introduction}\label{sec:Introduction}
\IEEEPARstart{T}{he} demand for online services including web search, online gaming, distributed file systems such as Google File System (GFS)~\cite{GFS}, and distributed Storage System such as BigTable~\cite{BigTable} and MapReduce~\cite{MapReduce} is growing exponentially. This significant growing demand for online services has led to a multitude of challenges in  Data Center Networks (DCNs) from DCN architecture design, congestion notification~\cite{zhang2013architecture,Conge1,Conge2,Conge3,Conge4,Conge5,Conge6,Conge7,Conge8}, TCP Incast~\cite{TCPIncast1,TCPIncast2,TCPIncast3}, virtual machine migration~\cite{VM1,VM2,VM3,VM4}, to routing in DCNs~\cite{Routing}.

Most importantly, due to the gravity of preparing DCNs as a scalable and reliable computing infrastructure,
online services run on hundreds of thousands of servers spread across large data centers have significantly craved electric power usage. Complying with such a growing demand in an environmentally friendly manner calls for innovations across different disciplines. Therefore, recently, studies on data centers have focused on reducing the energy consumption and accordingly the cost of electricity.
These studies can be largely categorized into two main approaches: power management techniques and green data centers. The first approach, which investigates CPU and memory power consumption, aims at reducing the carbon footprints and the cost of electricity.
The second approach, referred to as green data centers, not only tries to cut down the electricity consumption and its cost but also integrates renewable energy resources such as solar panels and wind farms into data centers, thereby promoting sustainability and green energy. In the past few years, a small and cohesive body of work investigated workload distribution across multiple data centers and the researchers came up with a variety of policies and algorithms. The social impacts of geographical load balancing is explored in~\cite{liu2011greening} and two distributed algorithms are provided that can be used to compute the optimal routing as well as provisioning decisions for Internet-scale systems. Another couple of research papers approach the problem by employing the mixed integer programming~\cite{rao2010minimizing,li2012towards}.
Also, Ghamkhari \textit{et al.}~\cite{ghamkhari2013energy} addressed the trade-off between minimizing a green
data center's energy costs and maximizing its revenue.
Also, Zhao \textit{et al.}~\cite{zhao2014dynamic} took into consideration of
dynamic VM pricing and designed a new algorithm to maximize the long-term cloud provider's profit. Moreover, Kiani and Ansari~\cite{kiani2015towards} proposed a workload distribution strategy based on the notion of green
workload and green service rate versus brown workload and brown service rate, respectively, and also real-time monitoring of the queue lengths.
%

In this paper, we propose a new workload distribution strategy for geographically dispersed green data centers in which our strategy aims at maximizing the revenue and minimizing the energy expenditures. To this end, we formulate an optimization framework for profit maximization which relies on the accuracy of the G/D/1 queue~\cite{ghamkhari2013energy,loss-shrof} in capturing the workload distribution. Moreover, our optimization-based workload distribution strategy taps on the merits of workload decomposition into green and brown workloads served by green and brown energy resources, respectively. In summary, we will address the following:
\begin{itemize}
\item {We develop a new model to maximize the profit of running geographically dispersed data centers. In our model, it is assumed that each data center is offering multiple classes of services and we take into account of individual SLA-deadline for each type of service. Also, we assume that each data center either has a renewable power source or is powered by a nearby wind or solar farm thereby taking into account of green energy. However, as the green energy resources may not be adequate to meet the QoS requirements for all incoming workloads, each data center is also provisioned by on-grid energy. Therefore, we further elaborate our model by taking into consideration of geographical electricity price diversity due to different electricity markets and time zones of the dispersed data centers.}
\item {Based on the developed model, we design an optimal workload distribution strategy in terms of the gained profit by the data centers. The profit is defined as $revenue-cost$ by considering the deadline, service income, penalty for the service
requests of each class, and also both green and brown energy costs. Our strategy relies on the accuracy of the G/D/1 queueing model in capturing the workload distribution. Furthermore, we prove the convexity of our optimization and therefore its appropriateness for practical purposes. In the optimization frameworks such as~\cite{ghamkhari2013energy} which are proposed for a single data center, the service rate is the only decision variable. However, as our model is an extension for a group of data centers, our objective function and the constraints are functions of both allocated workloads to the data centers and the service rate at each data center. In other words, we maximize the profit by not only optimizing the service rates at data centers but also allocating optimized workload to each data center. To prove the convexity of our problem, we introduce the average number of dropped requests
at each data center as an extended SLA constraint and based on that we can prove the convexity of the whole problem by using the convexity of the perspective of a function.}
\item{Our optimization model relies on the potential merit of the decomposition of the workload to the green and brown workload thereby taking into account of different costs and different environmental impacts of green and brown energy. In this way, we can allocate the green workload to the data centers based on the availability and cost variation of the green energy at different locations. However, for the brown workload, our strategy takes into account of electricity price diversity and hence
distinguishes the data centers by the price of electricity. In fact, we take into consideration of not only the cost of brown energy but also one time capital and maintenance expenses of renewable energy. Therefore, unlike some of the existing works in the literature, our optimal profit is not under the assumption that local renewable generation is always less than the local power consumption.}
\item{We evaluate the proposed workload distribution strategy via simulations and demonstrate that it outperforms the existing
workload distribution strategies in terms of the total profit.}
\end{itemize}
The rest of paper is organized as follows.
Sections~\ref{sec:System Model} and \ref{Problem Formulation}
describe the system model and problem formulation. In Sections~\ref{sec:Optimization Framework}, we propose our optimization framework.
Finally, Sections~\ref{sec:simultion} and \ref{sec:conclusion} present numerical results and conclude the paper, respectively.
\begin{figure}
\centering
\includegraphics[scale=.4]
{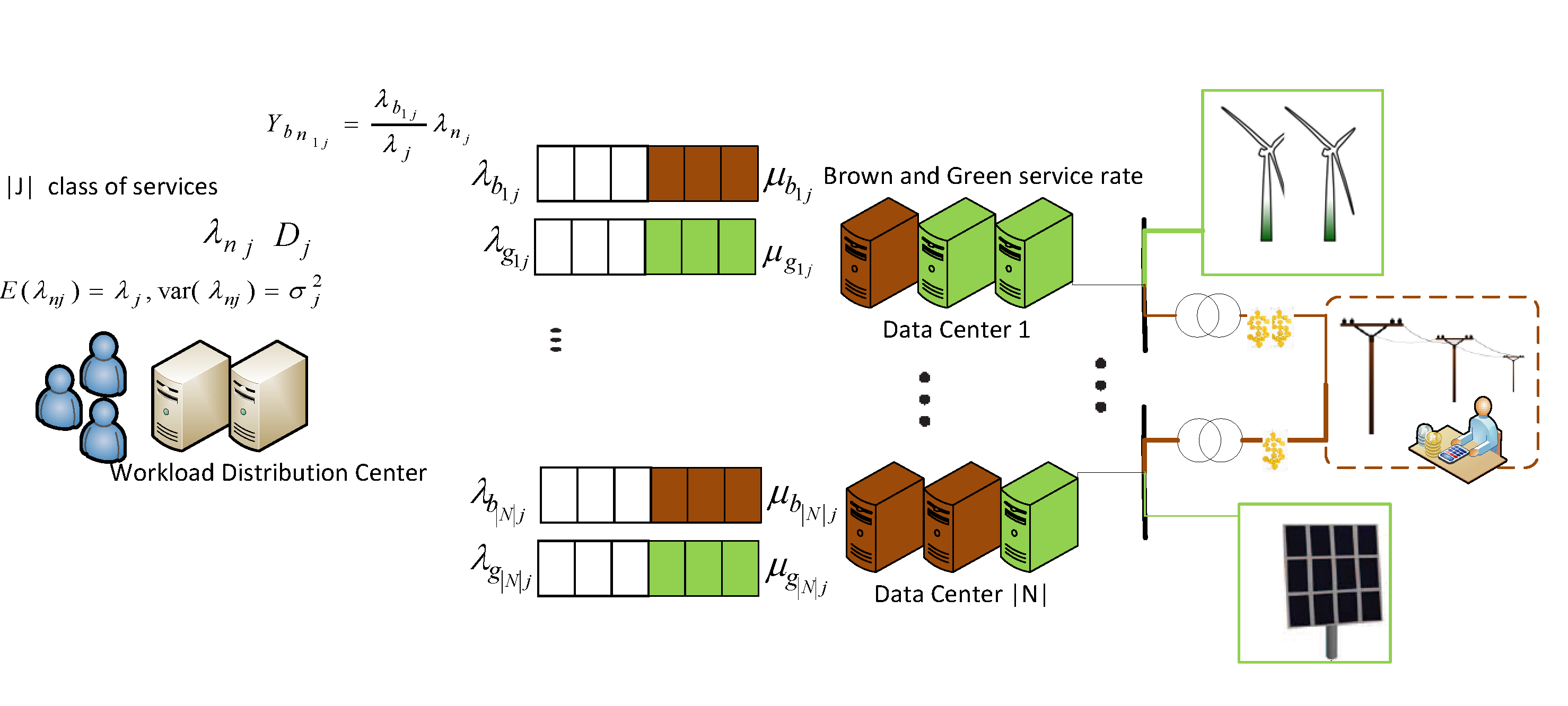}
\caption{System Model.}\label{fig:1}
\end{figure}

\section{System Model}\label{sec:System Model}
Figure~\ref{fig:1} shows the proposed system model in which we consider a group of $|N|$ data centers dispersed at different regions. Each data center is equipped with a collection of $M_i$ homogeneous servers.

The data centers are supplied by multiple types of power.
The major power supply of each data center is on-grid or brown energy. The data center has to pay brown energy prices according to its contract with the power company. The electricity pricing contract for each data center depends on the electricity markets at the data center's location. If the market is regulated, the electricity price has a flat rate during the day. On the other hand, if the region is following a deregulated market, the price of electricity is varying. In most cases, the data center pays less during off-peak hours and more during on-peak period. Therefore, we note the price variability among data centers located at different locations and time zones.

To reduce the cost of electricity and to capitalize on the environmental and sustainability advantages of green energy, we assume that each data center either is equipped with a renewable power source or has access to a nearby renewable energy source such as solar panels or a wind farm. It is worth mentioning that we assume the available renewable energy at each data center can only be used to supply power locally.

Each data center is offering $|J|$ multiple classes of service like web services, video streaming, etc. Each type of service has its specific deadline according to the SLA.

The service requests are initiated by users and arrive at the workload distribution center. One or a group of servers can serve as the workload distribution center~\cite{ghamkhari2012optimal}. These servers can be treated as the front-end devices that exist in multi-data center Internet services like Google and Itunes~\cite{Bianchini1}. The distribution center facilitates workload flexibility at the demand side. In other words, this center inspects the arriving requests from all users and manages the distribution
of the incoming workload to the geographically dispersed data centers based on the availability of green energy and the price of electricity.
In our formulation, the total power consumption at each data center takes into account of the Base Load and Proportional Load~\cite{ghamkhari2013energy},
\begin{eqnarray}\label{equ1}
Total~Power~Consumption~at~data~center~i=\nonumber\\
m_{i}[P_{idle}+(E_{usage}-1)P_{peak}]+m_{i}[(P_{peak}-P_{idle})U_{i}],
\end{eqnarray}
where the base load, $m_{i}[P_{idle}+(E_{usage}-1)P_{peak}]$, indicates the power consumption even when all of the turned on servers are idle.  The proportional load, $m_{i}[(P_{peak}-P_{idle})U_{i}]$, is the extra power consumption which is proportional to the CPU utilization of the servers, $U_{i}$, and accordingly to the workload. It is worth mentioning that both base and proportional loads are computed based on the number of switched on servers, $m_{i}$, idle power, $P_{idle}$, and average peak power of a single server, $P_{peak}$. Moreover, due to different energy efficiencies at different data centers, the definition of the total power consumption incorporates the Power Usage Effectiveness (PUE) ratio, $E_{usage}$, thereby amalgamating the power consumption at facility for cooling, lighting, and other overhead.~\cite{PUE}.

\section{Problem Formulation}\label{Problem Formulation}
We divide the running time of the data centers into a sequence of time slots at equal
length, $T$, e.g., a few minutes. Our goal is to maximize the data centers' total profit during the interval $T$. To this end, we propose an optimization problem to be solved at the beginning of each time slot in which we update the number of turned on servers as well as the allocated workload to each data center. Note that for the analysis, we consider a single time slot, e.g., $\Delta$ as the time slot of interest, and omit the explicit time dependence in the notations.

At the beginning of each time slot, we allocate the workload (total number of service requests) to the data centers based on the availability of green energy and the price of electricity. As the renewable energy and brown energy incur different
costs and different environmental impacts, we decompose the total workload into the green and brown workload. In fact, we distinguish the servers at each data center based on the energy which is utilized to power them. Some of the servers are turned on and powered by the available green energy (green servers), and the others, if needed, by purchasing brown energy (brown servers). Therefore, the distinction
between green and brown workloads is made mainly based on the server which is utilized to serve the workload. Specifically, the requests served by a green server are defined as the green workload and similarly those by a brown server the brown workload.

The data center's profit is modeled as $Revenue-Cost$, where the data center's revenue is calculated based on the QoS requirements satisfaction and the cost indicates the energy cost.
Owing to the limited computational resources at the data centers, the allocated requests to a data center are first placed in a queue before they can be processed by any available server. Accordingly, to satisfy the QoS requirements, the queueing delay for each service request should be limited by a deadline. If the data center can handle the service requests by the deadline, it receives the service income. Otherwise, it has to pay penalty to its customers. These three parameters, i.e., the deadline, service income, and penalty, depend on the type of service and are determined by the SLA~\cite{kusic2009power,ghamkhari2013energy}. Thus, we assume that the waiting requests of different classes of service at each data center are placed in different queues. Denote $D_j$, $\delta_j$, and $\gamma_j$ as the deadline, service income, and penalty for the service requests of class $j$, respectively. The service requests that are not handled by the deadlines are discarded~\cite{wilson2011better}.
In our problem formulation which is based on the workload decomposition, we distinguish the profit gained by serving green workload from the brown workload as the green and brown profit, respectively. To this end, we assume the green and brown requests of each class are placed in two different queues at a data center. In the next two subsections, we will formulate the green and brown profits.

\subsection{Green Profit Formulation}
We assume that the request rate of each class of service at the workload distribution center is a random process with an arbitrary and general probability distribution function, and $\lambda_{n_j}$ denotes the service request rate of class $j$ at time $n$.
Let $\lambda_j$ be the average rate of receiving service requests of class $j$ at the workload distribution center within time slot $\Delta$ of length $T$. Also, $\sigma^{2}_{j}$ denotes the variance of the class $j$ service request rate's probability distribution function.
Request interarrival times are assumed to be much shorter than a time slot duration, so that the request allocation can be based on the average arrival rate during the time slot.

We allocate $\frac{\lambda_{g_{ij}}}{\lambda_j}$ fraction of the service requests to the data center $i$'s green servers. These requests are first placed in a particular queue on green servers. The input process to this queue, i.e., $\lambda_{g_{n_{ij}}}=\frac{\lambda_{g_{ij}}}{\lambda_j}\lambda_{n_j}$, has the same general probability distribution function as the request rate of class $j$. Therefore, $\lambda_{g_{ij}}\neq0$ and $\sigma^{2}_{g_{ij}}=(\frac{\lambda_{g_{ij}}}{\lambda_j})^2\sigma^{2}_{j}$ are the mean and variance of the input process to the corresponding queue, respectively.

Based on the aforementioned QoS model, the green revenue earned by the data center $i$ for serving the green requests of different classes of service within a time slot can be calculated as,
$
R_{i}(\lambda_{g_{ij}},\mu_{g_{ij}})=
\sum_{j=1}^{|J|}([1-P_{L}(\lambda_{g_{ij}},\mu_{g_{ij}})]\delta_j\lambda_{g_{ij}}T-P_{L}(\lambda_{g_{ij}},\mu_{g_{ij}})\gamma_j\lambda_{g_{ij}}T)
$,
where $P_{L}(\lambda_{g_{ij}},\mu_{g_{ij}})$ denotes the probability that the waiting time for a service request of class $j$ exceeds its SLA-deadline. Note that $\mu_{g_{ij}}$ denotes the green service rate, i.e., the rate that the requests of class $j$ are removed (i.e., served) from the corresponding queue by the data center $i$'s green servers.

To obtain $P_{L}(\lambda_{g_{ij}},\mu_{g_{ij}})$, the SLA-deadline is translated into the loss probability of a G/D/1 queue.
In a nutshell, it is assumed the service rate that the service requests are removed from the queue, i.e., $\mu_{g_{ij}}$, is fixed over the time slot. Thus, for instance, if there are $Q_{ij}$ number of requests waiting in the queue upon the arrival of a new service request, it takes $\frac{Q_{ij}}{\mu_{g_{ij}}}$ seconds until the new request can be handled by any available server. If $\frac{Q_{ij}}{\mu_{g_{ij}}}\leq D_j$, then the new request can be handled before the deadline. Therefore, the SLA-deadline can be modeled by a finite-size queue with length $\mu_{g_{ij}}D_j$. In other words, in order to handle a new request by the SLA-deadline, it has to enter a queue with length $\mu_{g_{ij}}D_j$~\cite{ghamkhari2013energy}. According to queueing analysis~\cite{loss-shrof}, the loss probability of the finite-size queue can be accurately estimated from the tail of the queue length distribution for any general probability distribution. However, it is known that the estimation yields the highest level of accuracy when the service request rate is characterized by a Gaussian process~\cite{loss-shrof}. Therefore, through out the rest of this paper, the request rate of each class of service, accordingly the input process to the queues is assumed to be a Gaussian process, and the loss probability can be obtained as,

\begin{eqnarray}\label{equ3}
P_{L}(\lambda_{g_{ij}},\mu_{g_{ij}})=\alpha(\lambda_{g_{ij}},\mu_{g_{ij}})e^{-\frac{1}{2}\underset{n\geq1}{\min}M_{n}(\lambda_{g_{ij}},\mu_{g_{ij}})},
\end{eqnarray}
where

\begin{eqnarray}\label{equ4}
\alpha(\lambda_{g_{ij}},\mu_{g_{ij}})=~~~~~~~~~~~~~~~~~~~~~~~~~~~~~~~~~~~~~~~~~~~~~~~\nonumber\\
\frac{1}{\lambda_{g_{ij}}\sqrt{2\pi}\sigma_{g_{ij}}}e^\frac{(\mu_{g_{ij}}-\lambda_{g_{ij}})^2}{2\sigma_{g_{ij}}^2}\int_{\mu_{g_{ij}}}^{\infty}
(r-\mu_{g_{ij}})e^\frac{-(r-\lambda_{g_{ij}})^2}{2\sigma_{g_{ij}}^2}dr,&
\end{eqnarray}
and for each $n\geq1$,
\begin{eqnarray}\label{equ5}
M_{n}(\lambda_{g_{ij}},\mu_{g_{ij}})=\frac{((D_j-d_i)\mu_{g_{ij}}+n(\mu_{g_{ij}}-\lambda_{g_{ij}}))^2}
{nC_{\lambda_{g_{ij}}}(0)+2\sum_{l=1}^{n-1}(n-l)C_{\lambda_{g_{ij}}}(l)},
\end{eqnarray}
where $C_{\lambda_{g_{ij}}}(l)$ is the autocovariance of the class $j$ service request rate's probability function at data center $i$, and we have $\sigma^{2}_{g_{ij}}=C_{\lambda_{g_{ij}}}(0)$. Also, $d_i$ is the network delay experienced by a request from the workload distribution center to data center $i$.

The green power consumption at each data center depends on the number of switched on green servers as well as the CPU utilization of each green server. The total number of switched on green servers at data center $i$ can be expressed based on the total green service rate as
$m_{g_i}=\sum_{j=1}^{|J|}\frac{\mu_{g_{ij}}}{k_j}$,
where each server can handle $k_j$ service requests of class $j$ per second.
Also, within the interval of $T$, each switched on green server handles
$\frac{T(1-P_{L}(\lambda_{g_{ij}},\mu_{g_{ij}}))\lambda_{g_{ij}}}{m_{g_i}}$
requests of class $j$~\cite{ghamkhari2013energy}.
Thus, the total CPU busy time of each server can be obtained as
$\sum_{j=1}^{|J|}\frac{T(1-P_{L}(\lambda_{g_{ij}},\mu_{g_{ij}}))\lambda_{g_{ij}}}{m_{g_i}k_j}$.
By dividing the total server busy time by $T$, we have the CPU utilization
$U_{g_i}=\sum_{j=1}^{|J|}\frac{(1-P_{L}(\lambda_{g_{ij}},\mu_{g_{ij}}))\lambda_{g_{ij}}}{m_{g_i}k_j}$.
Therefore, referring to the definition of power consumption in~(\ref{equ1}), the total green power consumption in data center $i$ at the time of interest can be expressed as,

\begin{eqnarray}\label{equ10}
E_{i}(\lambda_{g_{ij}},\mu_{g_{ij}})=(P_{idle}+(E_{usage}-1)P_{peak})\sum_{j=1}^{|J|}\frac{\mu_{g_{ij}}}{k_j}+\nonumber\\
(P_{peak}-P_{idle})\sum_{j=1}^{|J|}\frac{(1-P_{L}(\lambda_{g_{ij}},\mu_{g_{ij}}))\lambda_{g_{ij}}}{k_j}.
\end{eqnarray}
Note that the total number of the green servers at each data center, and accordingly the green service rate is limited by the available green energy at the time slot of interest. Let $W_i$ be the available green energy at data center $i$ within the time slot. $W_i$ is predicted at the beginning of the time slot, and depends, for example, on wind speed and solar irradiance.
Similar to some other published papers in the literature such as~\cite{ghamkhari2012optimal,ghamkhari2013energy} it is assumed that the time slot is small enough (e.g., every few minutes). Therefore, while the amount of renewable energy is changing at different time of a day, it is reasonable that solar irradiance and wind speed are relatively stable within a slot.
We assume $C_{g_i}$ is the cost of renewable energy at data center $i$. The cost of green energy generation includes one time capital and maintenance expenses.
The average unit cost of renewable energy can be obtained by averaging over the total amount of energy generated during the whole operation period.
Therefore, the total green profit gained by all the data centers during the time slot of interest can be calculated as
$
Profit_g=\sum_{i=1}^{|N|}(R_{i}(\lambda_{g_{ij}},\mu_{g_{ij}})-C_{g_i}TE_{i}(\lambda_{g_{ij}},\mu_{g_{ij}}))
$.

\subsection{Brown profit formulation}
If green energy generation is not adequate to serve all incoming workload, brown energy is purchased. Brown energy is considered as an additional resource to power on additional servers referred to as the brown servers. We allocate
$\lambda_{b_{n_{ij}}}=\frac{\lambda_{b_{ij}}}{\lambda_j}\lambda_{n_j}$ service requests, as the brown requests, to the data center $i$'s brown servers. These requests are first placed in their particular queue on brown servers, and $\lambda_{b_{ij}}\neq0$ and $\sigma^{2}_{b_{ij}}=(\frac{\lambda_{b_{ij}}}{\lambda_j})^2\sigma^{2}_{j}$ are the mean and variance of the input process to the queue, respectively. 
When using brown energy, we note the different deregulated electricity markets of data centers located at different regions. Denote $C_{b_i}$ as the price of electricity at data center $i$ within the time slot of interest. In order to benefit from the electricity price diversity, the distribution center can employ the day-ahead electricity price forecasting methods~\cite{wu2010hybrid,areekul2010hybrid}. Therefore, the total brown profit gained by all the data centers during the time slot of interest can be calculated as,
$
Profit_b=\sum_{i=1}^{|N|}(R_{b_i}(\lambda_{b_{ij}},\mu_{b_{ij}})-C_{b_i}TE_{b_i}(\lambda_{b_{ij}},\mu_{b_{ij}})).
$
In the next section, we propose an optimization framework for the service request distribution. The objective of our framework is to maximize the total profit earned by the data centers within each time slot. Our optimization framework uses the results of renewable energy and electricity price forecasting methods.

\section{Optimization Framework}\label{sec:Optimization Framework}
In order to maximize the total profit earned by the data centers, we update the allocated workload and the service rates for each data center. In fact, we seek to maximize the total profit by optimizing the allocated green and brown requests (i.e., $\lambda_{g_{ij}}$ and $\lambda_{b_{ij}}$) as well as the green and brown service rates (i.e., $\mu_{g_{ij}}$ and $\mu_{b_{ij}}$) within each time slot. To this end, the following optimization problem is proposed to be solved at the beginning of the time slot of interest,
\begin{equation}\label{equ18}
\underset{\lambda_{g_{ij}},\mu_{g_{ij}},\lambda_{b_{ij}},\mu_{b_{ij}}}{\text{maximize}}~~
(Profit_g+Profit_b)~~~~~~~~~~~~~~~~~~~~~~~~~
\end{equation}

\begin{equation}
\text{subject to}~~\nonumber
\end{equation}

\begin{equation}\label{equ19}
\begin{aligned}
0<\lambda_{g_{ij}}\leq\mu_{g_{ij}},~\forall i\in N,~\forall j\in J,
\end{aligned}
\end{equation}

\begin{equation}\label{equ20}
\begin{aligned}
0<\lambda_{b_{ij}}\leq\mu_{b_{ij}},~\forall i\in N,~\forall j\in J,
\end{aligned}
\end{equation}

\begin{equation}\label{equ21}
\begin{aligned}
\sum_{j=1}^{|J|}\frac{\mu_{g_{ij}}}{k_j}\leq\lfloor\frac{W_{i}(t)}{P_{peak}E_{usage}}\rfloor,~\forall i\in N,
\end{aligned}
\end{equation}

\begin{equation}\label{equ22}
\begin{aligned}
\sum_{i=1}^{|N|}(\lambda_{g_{ij}}+\lambda_{b_{ij}})=\lambda_j,~\forall j\in J,
\end{aligned}
\end{equation}

\begin{equation}\label{equ23}
\begin{aligned}
\lambda_{g_{ij}}P_{L}(\lambda_{g_{ij}},\mu_{g_{ij}})\leq~TH_j,~\forall i\in N,~\forall j\in J,
\end{aligned}
\end{equation}
\begin{equation}\label{equ24}
\begin{aligned}
\lambda_{b_{ij}}P_{L}(\lambda_{b_{ij}},\mu_{b_{ij}})\leq~TH_j,~\forall i\in N,~\forall j\in J,
\end{aligned}
\end{equation}
where the inequality constraints~(\ref{equ19}), (\ref{equ20}) are to lower bound the service rate of each queue by the average of the input process to that queue and are necessary for stabilizing the service request queue. In addition, the inequality constraint~(\ref{equ21}) is used to limit the green service rates by the available renewable energy in which we make full CPU utilization assumption. Also, we use equality constraint~(\ref{equ22}) to allot all the requests of each class to the data centers based on the average rate of receiving service requests. Moreover, by inequality constraints~(\ref{equ23}), (\ref{equ24}), we add an extended SLA requirement in which the average number of dropped requests at each queue is upper bounded by a constant $TH_j$.

The proposed optimization problem is a convex optimization problem, as proven in the following theorem, and consequently can be solved by efficient optimization techniques, such as the interior point method (IPM).
\begin{theorem}\label{theorem1}
\label{T1}
The constrained optimization problem~(\ref{equ18}) is a convex optimization problem if data centers are profitable for each class of service and
\begin{equation}\label{equ25}
\begin{aligned}
\mu_{g_{ij}}\geq1~~and~~\mu_{b_{ij}}\geq1, \forall~i,j
\end{aligned}
\end{equation}
\end{theorem}


\begin{figure}[ht]
\begin{minipage}[b]{0.53\linewidth}
\includegraphics[scale=.27]
{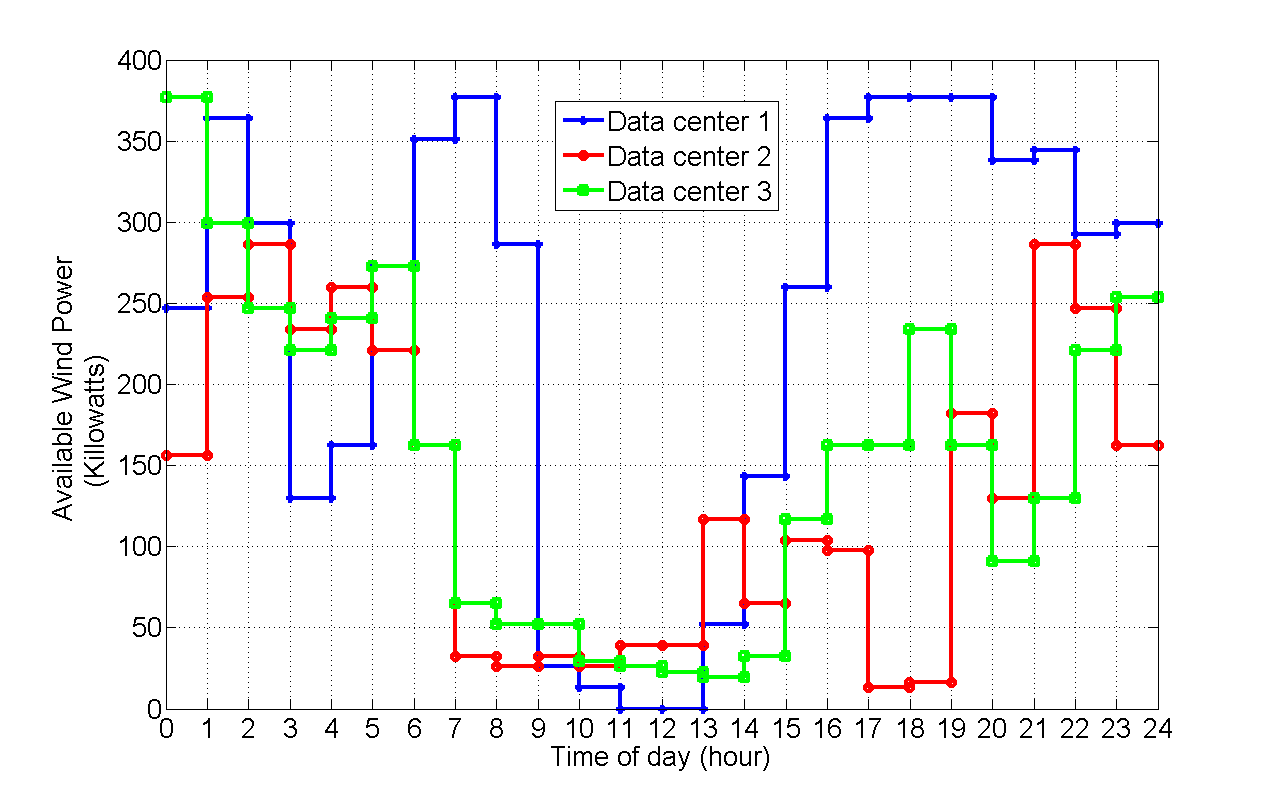}
\caption{Wind power generation.}
\label{fig:2}
\end{minipage}
\begin{minipage}[b]{0.53\linewidth}
\includegraphics[scale=.27]
{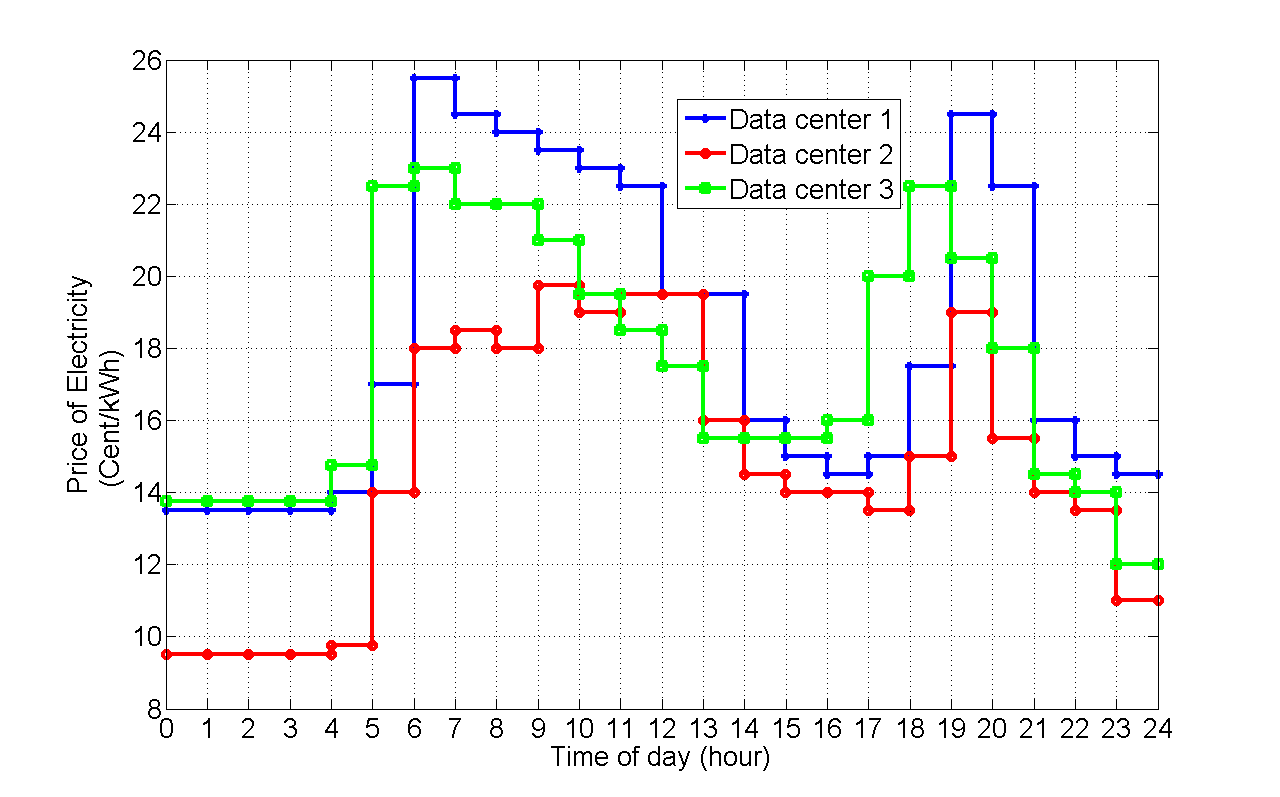}
\caption{Price of electricity.}
\label{fig:3}
\end{minipage}
\end{figure}
\begin{figure}[ht]
\begin{minipage}[b]{0.53\linewidth}
\includegraphics[scale=.27]
{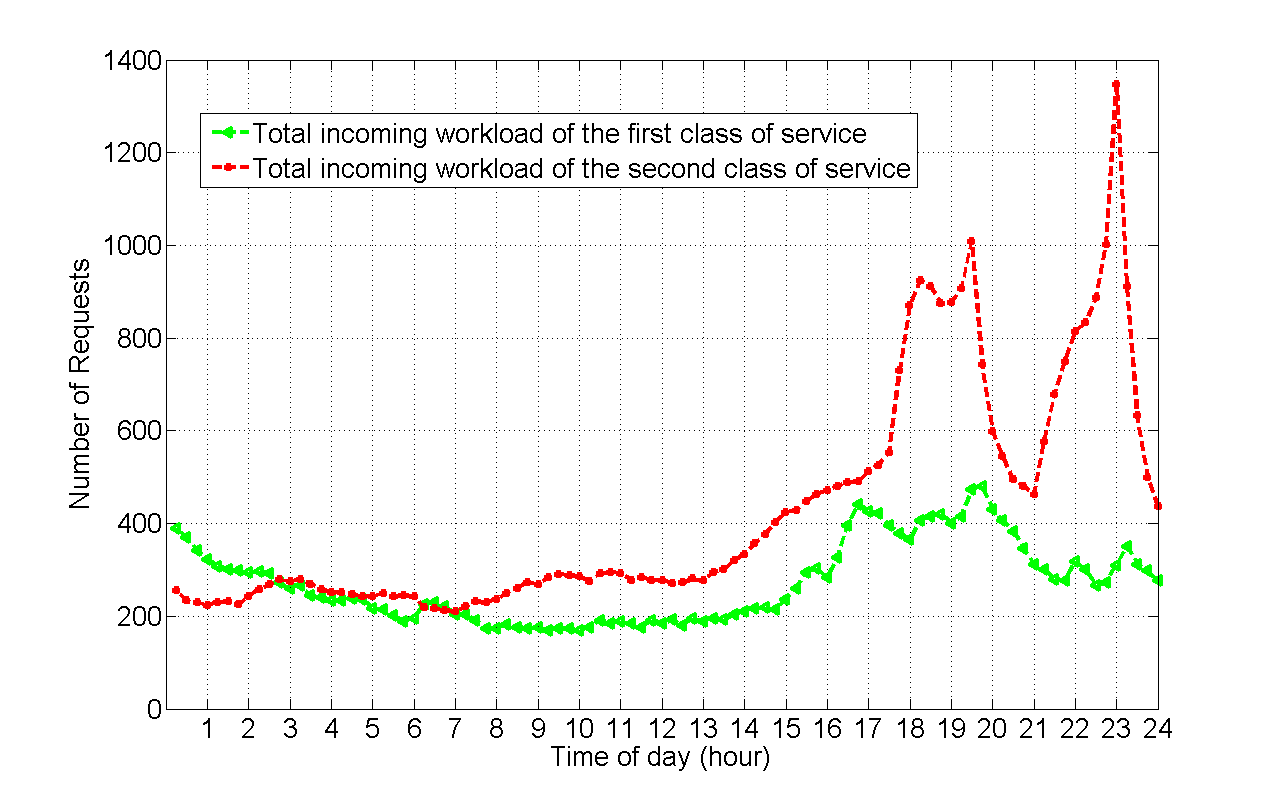}
\caption{The total incoming workload.}
\label{fig:4}
\end{minipage}
\begin{minipage}[b]{0.53\linewidth}
\includegraphics[scale=.27]
{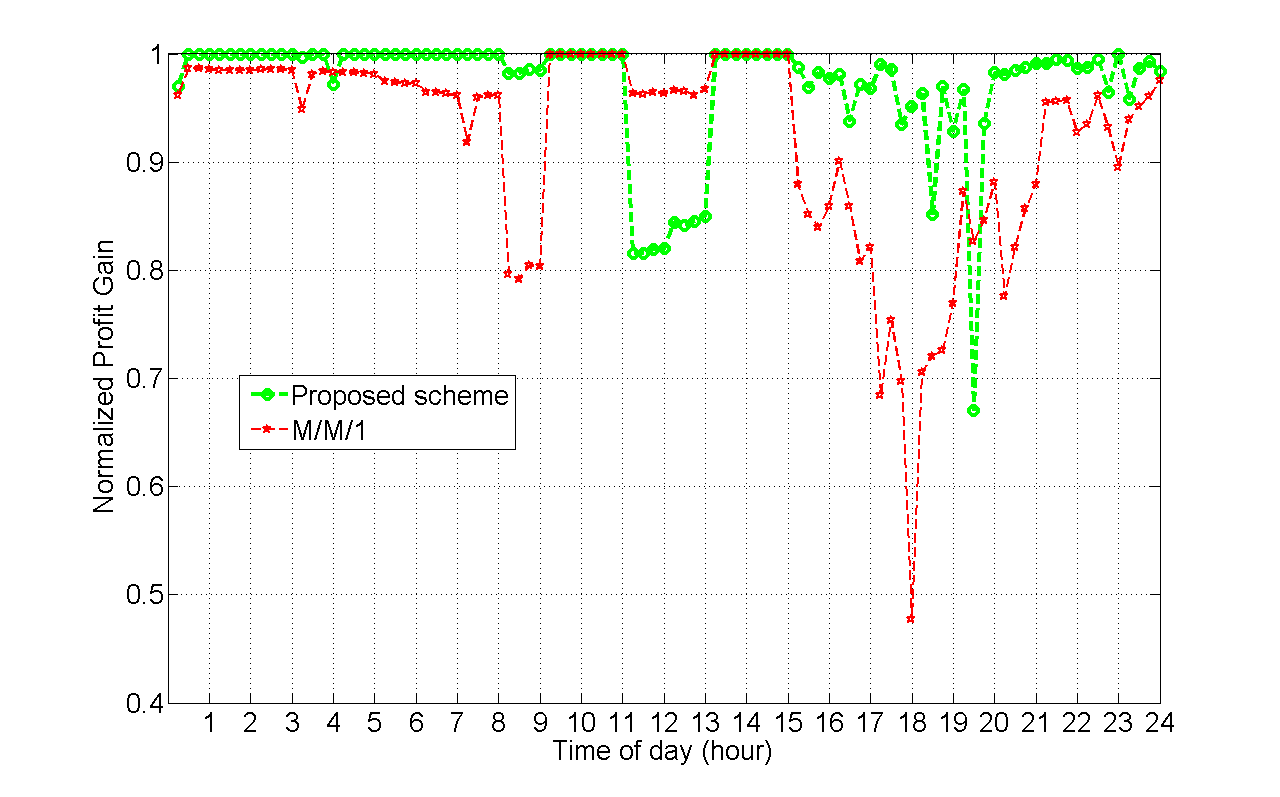}
\caption{Normalized profit gain.}
\label{fig:5}
\end{minipage}
\end{figure}
The proof of Theorem~\ref{T1} is given in the appendix. It is worth mentioning that the G/D/1 model in~\cite{loss-shrof} is valid only for the range of service rates, $\mu_{g_{ij}}\geq\lambda_{g_{ij}}$ and $\mu_{b_{ij}}\geq\lambda_{g_{ij}}$, which we have already considered in our constraints. Therefore, even if we do not allocate any workload to a data center, the service rate has to be set greater than one for the problem to be convex.
\vspace{-.1in}
\section{Simulation Results}\label{sec:simultion}
We consider $|N| = 3$ data centers offering $|J|=2$ different classes of service. Each data center is integrated with a wind farm as a renewable power source. It is assumed that the data centers are located at three different regions with deregulated electricity market. Our simulation data are based on the trends of wind power and electricity price shown in Figures~\ref{fig:2} and~\ref{fig:3}, respectively, which are updated every hour. We simulated the total workload of two classes of service using two sample days of the requests made to the 1998 World Cup web site~\cite{Worldcup} shown in Figure~\ref{fig:4}. Also, for each turned on server, we have assumed $P_{peak} = 0.2$ kw, $P_{idle} = 0.1$ kw, and $E_{usage} = 1.2$.
\begin{figure}[ht]
$\begin{array}{ll}
\includegraphics[scale=.27]
{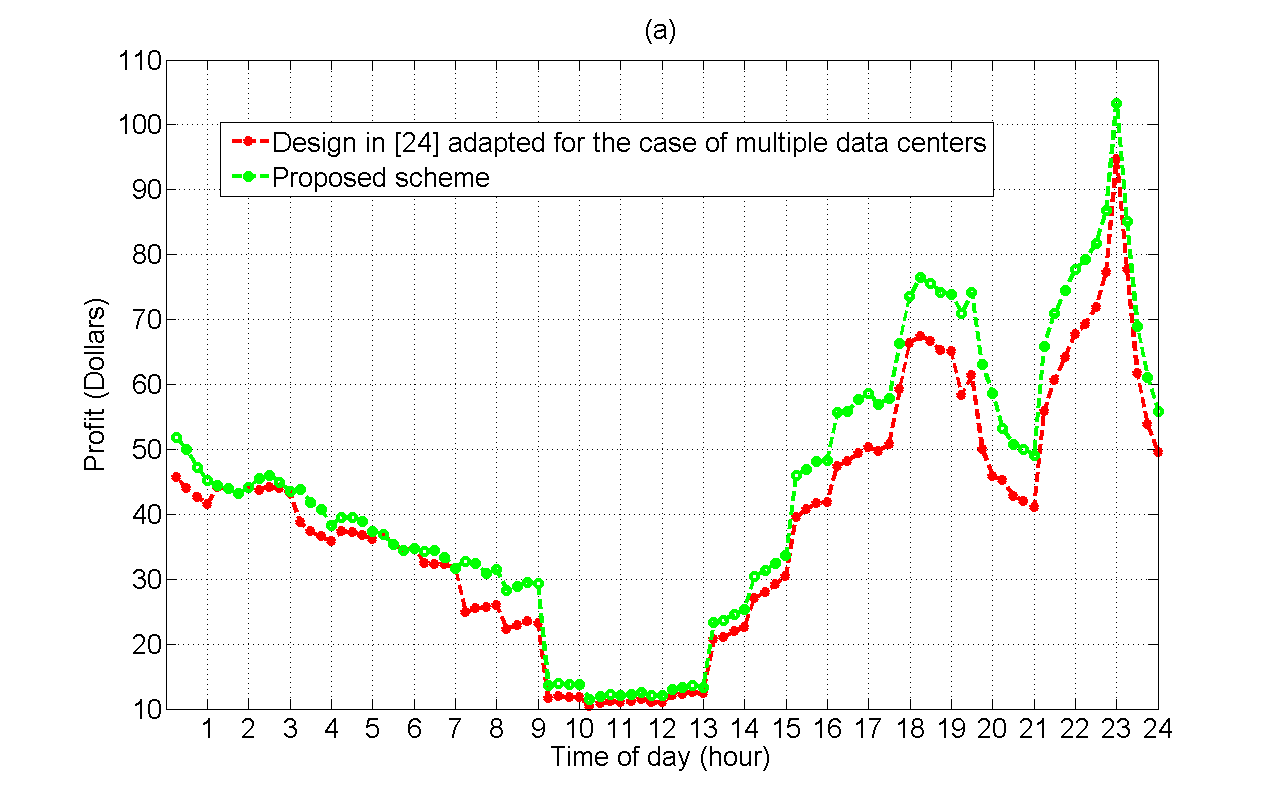}
\includegraphics[scale=.27]
{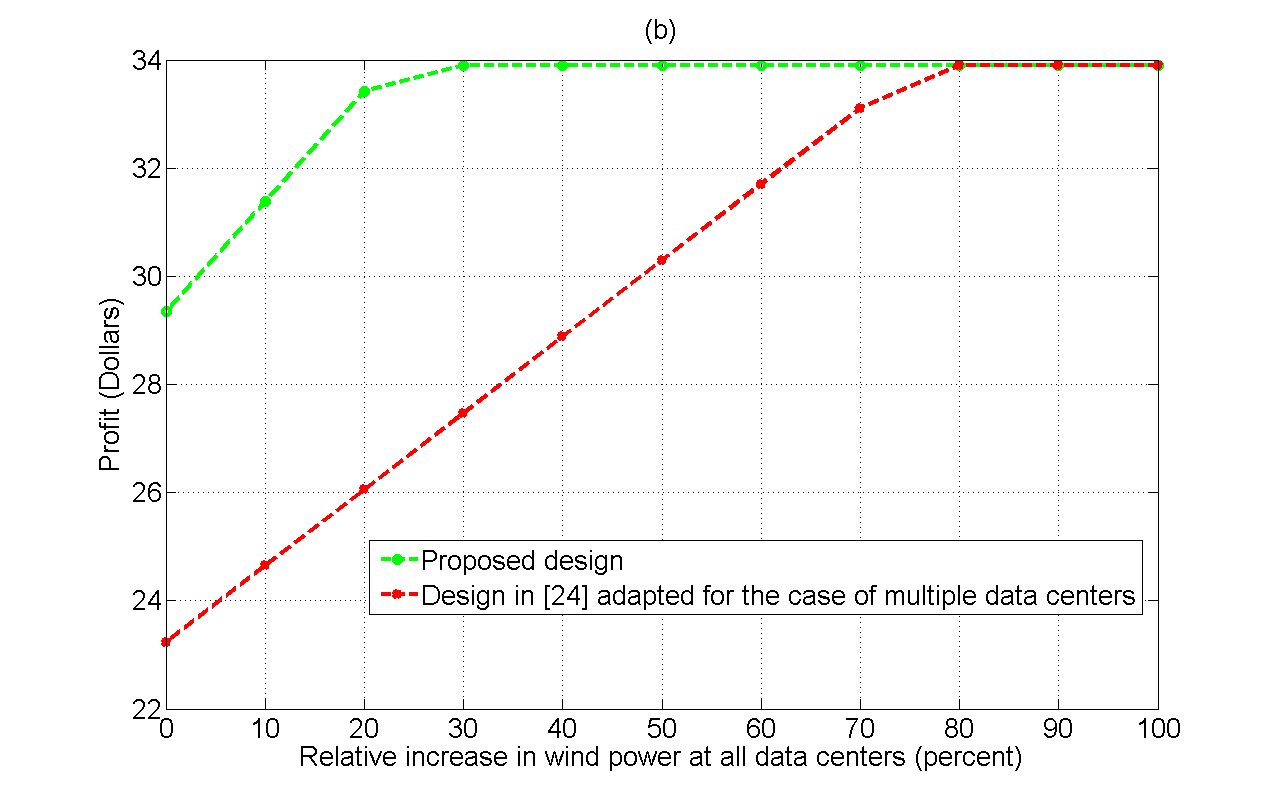}
\end{array}$
\caption{Performance comparison between the profit gain of the proposed design and design in~\cite{ghamkhari2013energy} adopted for the case of multiple data centers. (a) 24 hours operation. (b) One time slot. }
\label{fig:6}
\end{figure}
\begin{figure}[ht]
\begin{center}
$\begin{array}{ll}
\includegraphics[scale=0.27]{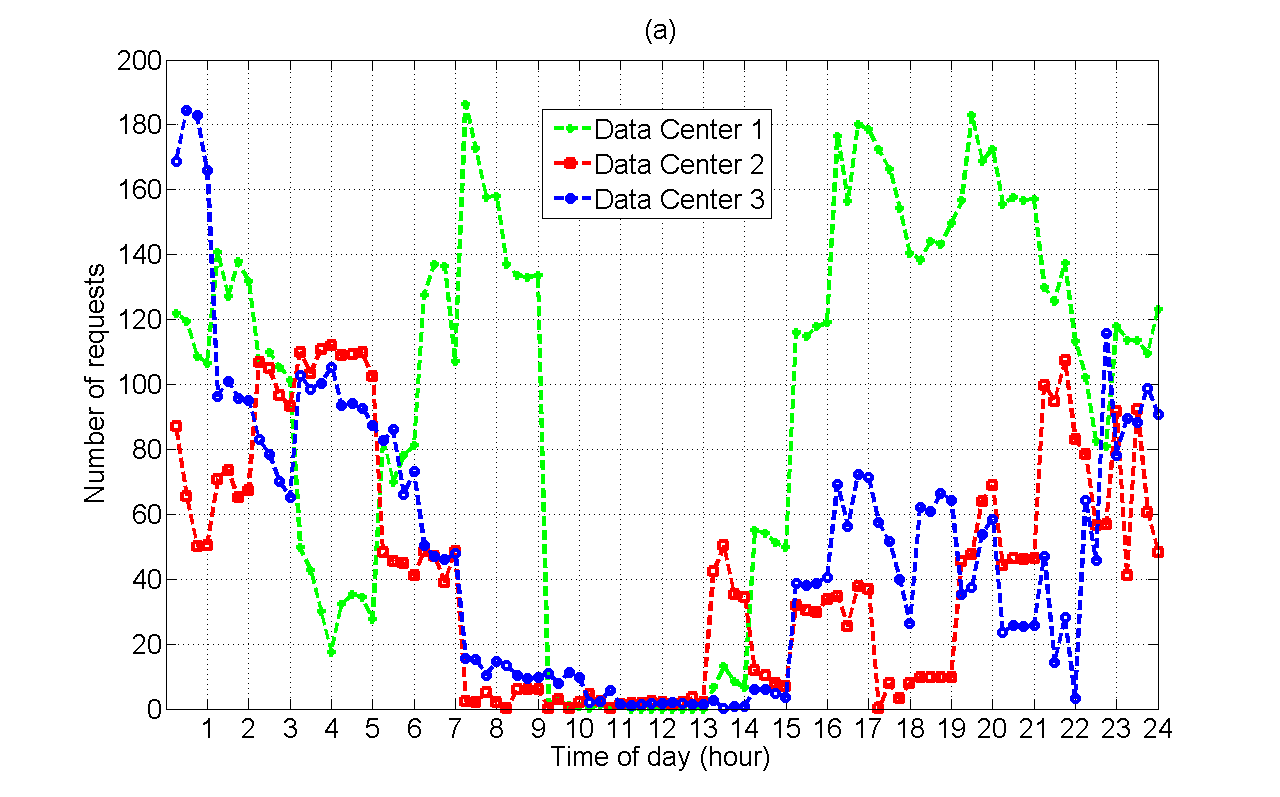}
\includegraphics[scale=0.27]{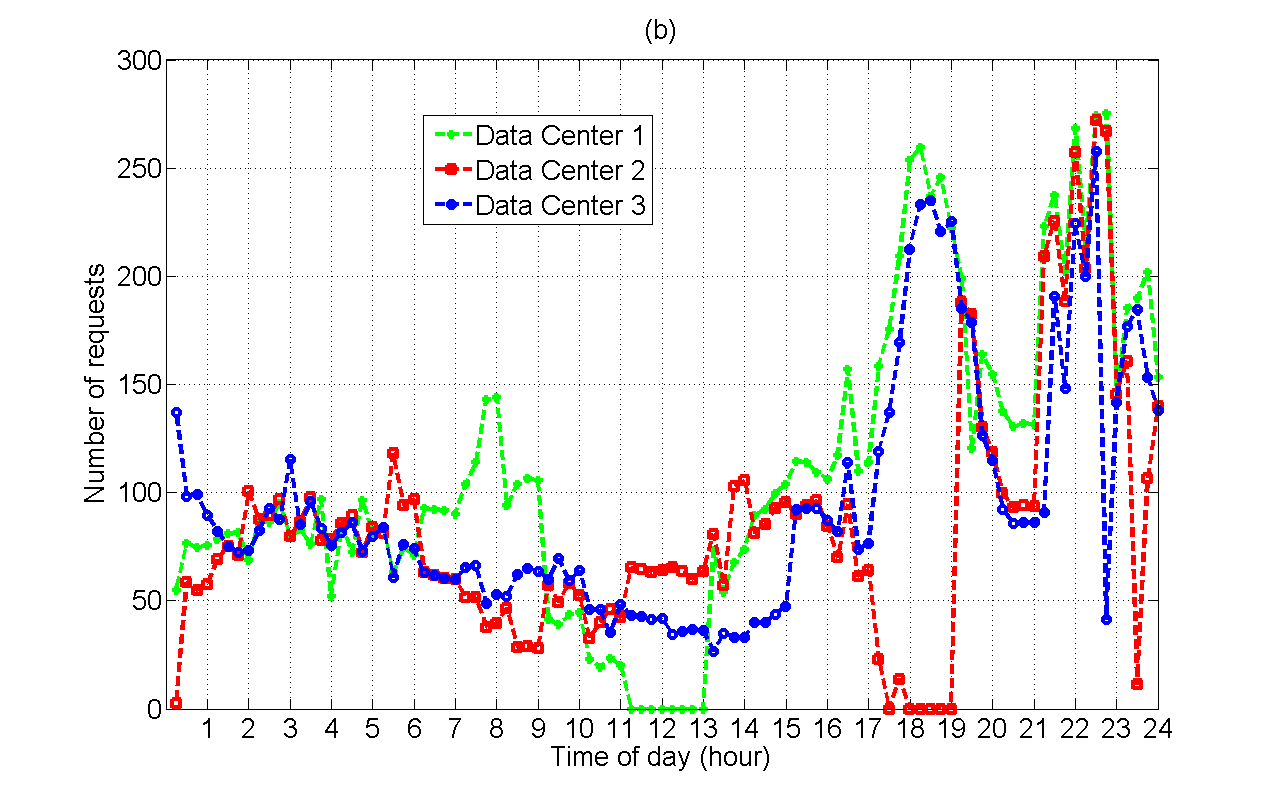}
\end{array}$
\end{center}
\caption{Allocated green workload to the data centers. (a) First class of service. (b) Second class of service. }
\label{fig:7}
\end{figure}
\begin{figure}[ht]
\begin{center}
$\begin{array}{ll}
\includegraphics[scale=0.27]{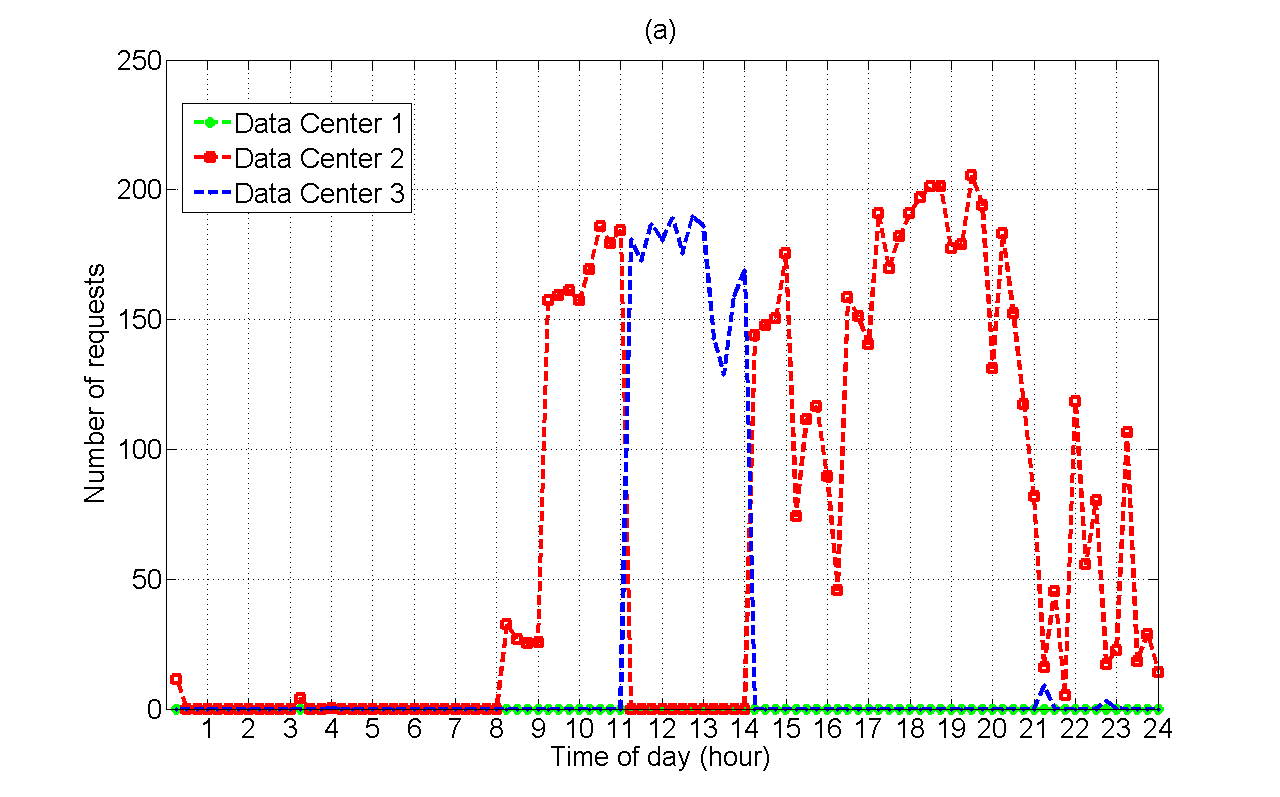}
\includegraphics[scale=0.27]{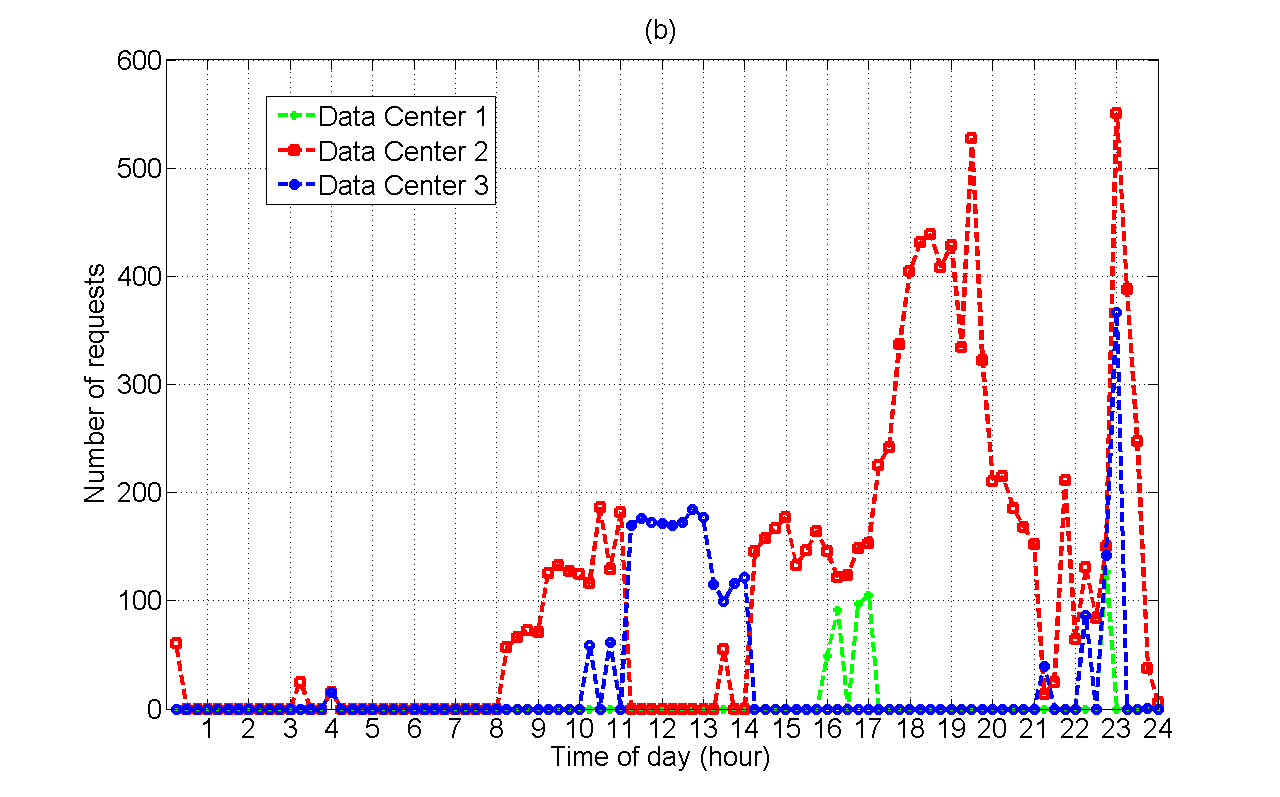}
\end{array}$
\end{center}
\caption{Allocated brown workload to the data centers. (a) First class of service. (b) Second class of service. }
\label{fig:8}
\end{figure}




Figure~\ref{fig:5} compares the normalized profit gained by running three data centers. As shown in this figure, the curves represent the normalized profit of our proposed optimization problem and the design which is based on M/M/1 queueing~\cite{rao2012coordinated}.
The normalized profit gain is calculated as $(Profit-Profit_{Base})/(Profit_{Max}-Profit_{Base})$ where $Profit_{Base}$ is the profit obtained when $\mu=\lambda$ and $Profit_{Max}$ is the maximum of the profit curve obtained by simulation~\cite{ghamkhari2013energy}.
We can see that the proposed design outperforms the normalized profit gain of M/M/1 queueing because the G/D/1 queueing model can capture the workload distribution more accurately than M/M/1.
Also, Figure~\ref{fig:6} demonstrates the better performance of our proposed design than the design in~\cite{ghamkhari2013energy} adapted for the case of multiple data centers. While Figure~\ref{fig:6}(a) compares the gained profits of 24 hours operation of the data
centers, Figure~\ref{fig:6}(b) shows the gained profit of a sample time slot versus the relative increase in green energy.
Figure~\ref{fig:7}(a) and (b) demonstrate the allocated green workloads of the first and second class of service to each data center, respectively. For example, the trend of wind power indicates that after hour 15 most of the green workload is assigned
to data center 1 where the highest wind power is available. However, from hours 10 to 13, the available wind power at data center 1 is lower than the other data centers, and thus less of the green workload is allocated to this data center.
Finally, Figure~\ref{fig:8} shows the allocated brown workloads of the first and second class of
service to each data center. For example, as shown in the Figure~\ref{fig:8}(a), from hours 8 to 11, all of the left over of the requests of both classes (the requests that are not served by green energy) are allocated to data center 2 where the price of electricity is the lowest.

\vspace{-.1in}
\section{Conclusion}\label{sec:conclusion}
In this paper, we have developed a new model to maximize the profit of running geographically dispersed data centers. Our model considers multiple classes of service and takes into account of individual SLA-deadline for each type of service. Furthermore, our model is elaborated by taking into consideration of geographical electricity price diversity due to different electricity markets at each data center's location and the availability of renewable energy.

Based on the developed model, we have designed an optimization-based workload distribution scheme that relies on the accuracy of G/D/1 queue in characterizing the workload distribution and  the workload decomposition to the green and brown workload. We have also proven the convexity of the formulated optimization problem and evaluated the performance of our workload distribution scheme via extensive simulations.
\vspace{-.1in}
\bibliographystyle{IEEE}
\bibliography{refs}
\vspace{-.1in}
\appendix
\begin{IEEEproof}[Proof of Theorem \ref{theorem1}]
To show the convexity of the proposed optimization problem, we require to prove~\cite{boyd2009convex}:
\begin{itemize}
\item{The objective function, i.e., $Profit_g+Profit_b$, is concave.}
\item{The inequality constraint functions are convex.}
\item{The equality constraint functions, i.e., $\sum_{i=1}^{|N|}(\lambda_{g_{ij}}+\lambda_{b_{ij}})-\lambda_j$, are affine.}
\end{itemize}
Since the corresponding functions of the constraints (\ref{equ19}), (\ref{equ20}), (\ref{equ21}) and (\ref{equ22}) are all linear, we start by proving the convexity of the following function,
\begin{equation}\label{equ26}
\begin{aligned}
f(\lambda_{g_{ij}},\mu_{g_{ij}})\triangleq\lambda_{g_{ij}}P_{L}(\lambda_{g_{ij}},\mu_{g_{ij}})-TH_j,~\forall i\in |N|,~\forall j\in |J|.
\end{aligned}
\end{equation}
From~(\ref{equ3}), as $e^{-x}$ is non-increasing, we have
\begin{equation}\label{equ27}
P_{L}(\lambda_{g_{ij}},\mu_{g_{ij}})=\underset{n\geq1}{\max} ~\alpha(\lambda_{g_{ij}},\mu_{g_{ij}})e^{-\frac{1}{2}M_{n}(\lambda_{g_{ij}}, \mu_{g_{ij}})}.
\end{equation}
Since $\max$ preserves convexity~\cite{boyd2009convex} and $TH_j$ is constant, the function $f(\lambda_{g_{ij}},\mu_{g_{ij}})$ is proven to be convex if we can prove the following function,
\begin{equation}\label{equ28}
f_n(\lambda_{g_{ij}},\mu_{g_{ij}})=\lambda_{g_{ij}} \alpha(\lambda_{g_{ij}},\mu_{g_{ij}})e^{-\frac{1}{2}M_{n}(\lambda_{g_{ij}}, \mu_{g_{ij}})},
\end{equation}
is convex for each $n\geq1$.

After reordering the terms in~(\ref{equ4}), we can show that,
\begin{equation}\label{equ29}
\begin{aligned}
\alpha(\lambda_{g_{ij}},\mu_{g_{ij}})=~~~~~~~~~~~~~~~~~~~~~~~~~~~~~~~~~~~~~~~~~~~~~~~~~~~~~~~~\\
\frac{\sigma_{g_{ij}}}{\lambda_{g_{ij}}\sqrt{2\pi}}[1-\frac{(\mu_{g_{ij}}-\lambda_{g_{ij}})}{\sigma_{g_{ij}}}
e^\frac{(\mu_{g_{ij}}-\lambda_{g_{ij}})^2}{2\sigma_{g_{ij}}^2}\int_{\frac{(\mu_{g_{ij}}-\lambda_{g_{ij}})}{\sigma_{g_{ij}}}}^{\infty}
e^\frac{-u^2}{2}du]\\
\end{aligned}
\end{equation}
By substituting $\sigma_{g_{ij}}=(\frac{\lambda_{g_{ij}}}{\lambda_j})\sigma_{j}$ in~(\ref{equ29}) and (\ref{equ5}) respectively, and after simple algebraic manipulation we have,
\begin{equation}\label{equ30}
\begin{aligned}
\alpha(\lambda_{g_{ij}},\mu_{g_{ij}})=~~~~~~~~~~~~~~~~~~~~~~~~~~~~~~~~~~~~~~~~~~~~~~~~~~~~~~~~\\
\frac{C_{v_j}}{\sqrt{2\pi}}[1-\frac{1}{C_{v_j}}(\frac{\mu_{g_{ij}}}{\lambda_{g_{ij}}}-1)
e^{\frac{1}{2C_{v_j}^2}(\frac{\mu_{g_{ij}}}{\lambda_{g_{ij}}}-1)^2}\int_{\frac{1}{C_{v_j}}(\frac{\mu_{g_{ij}}}{\lambda_{g_{ij}}}-1)}^{\infty}
e^\frac{-u^2}{2}du]\\
\end{aligned}
\end{equation}
and
\vspace{-.1in}
\begin{equation}\label{equ31}
\begin{aligned}
M_{n}(\lambda_{g_{ij}},\mu_{g_{ij}})=\frac{((D_j-d_i+n)(\frac{\mu_{g_{ij}}}{\lambda_{g_{ij}}}-1)+(D_j-d_i))^2}
{\rho_{n_j}},
\end{aligned}
\end{equation}
where $C_{v_j}=\frac{\sigma_j}{\lambda_j}$ is the coefficient of variation of the class $j$'s service request rate. Also,
\vspace{-.1in}
\begin{equation}\label{equ32}
\begin{aligned}
\rho_{n_j}\triangleq~nC_{v_j}^2+2\sum_{l=1}^{n-1}(n-l)\frac{C_{\lambda_{j}}(l)}{\lambda_j^2}
\end{aligned}
\end{equation}
Equations (\ref{equ30}) and (\ref{equ31}) indicate that $f_n(\lambda_{g_{ij}},\mu_{g_{ij}})$ is the perspective of the following function,
\begin{equation}\label{equ33}
\begin{aligned}
g_n(\mu_{g_{ij}})=\alpha(\mu_{g_{ij}})e^{-\frac{1}{2}M_{n}(\mu_{g_{ij}})},
\end{aligned}
\end{equation}
where
\vspace{-.1in}
\begin{equation}\label{equ34}
\begin{aligned}
\alpha(\mu_{g_{ij}})=~~~~~~~~~~~~~~~~~~~~~~~~~~~~~~~~~~~~~~~~~~~~~~~~~~~~~~~~~~~\\
\frac{C_{v_j}}{\sqrt{2\pi}}[1-\frac{1}{C_{v_j}}(\mu_{g_{ij}}-1)
e^{\frac{1}{2C_{v_j}^2}(\mu_{g_{ij}}-1)^2}\int_{\frac{1}{C_{v_j}}(\mu_{g_{ij}}-1)}^{\infty}
e^\frac{-u^2}{2}du]
\end{aligned}
\end{equation}
and
\vspace{-.1in}
\begin{equation}\label{equ35}
\begin{aligned}
M_{n}(\mu_{g_{ij}})=\frac{((D_j-d_i+n)(\mu_{g_{ij}}-1)+(D_j-d_i))^2}
{\rho_{n_j}}.
\end{aligned}
\end{equation}
If $g_n(\mu_{g_{ij}})$ is convex, so is its perspective function $f_n(\lambda_{g_{ij}},\mu_{g_{ij}})$~\cite{boyd2009convex}. Therefore, we continue our proof by proving the convexity of $g_n(\mu_{g_{ij}})$. Let's define
\vspace{-.1in}
\begin{equation}\label{equ36}
\begin{aligned}
t\triangleq\frac{(\mu_{g_{ij}}-1)}{C_{v_j}}
\end{aligned}
\end{equation}
Then, we have
\vspace{-.13in}
\begin{equation}\label{equ37}
\begin{aligned}
g_n(t)=\alpha(t)e^{-\frac{1}{2}M_{n}(t)}
\end{aligned}
\end{equation}
\begin{equation}\label{equ38}
\begin{aligned}
\alpha(t)=
\frac{C_{v_j}}{\sqrt{2\pi}}[1-t
e^{\frac{t^2}{2}}\int_{t}^{\infty}
e^\frac{-u^2}{2}du]
\end{aligned}
\end{equation}
and
\vspace{-.13in}
\begin{equation}\label{equ39}
\begin{aligned}
M_{n}(t)=\frac{((D_j-d_i+n)C_{v_j}t+(D_j-d_i))^2}
{\rho_{n_j}}.
\end{aligned}
\end{equation}
Then, the function $g_n(\mu_{g_{ij}})$ is proven to be convex if we can show for each $n\geq1$,
\vspace{-.13in}
\begin{equation}\label{equ40}
\begin{aligned}
g^{\prime\prime}_n(t)=e^{-\frac{1}{2}M_{n}(t)}(\alpha^{\prime\prime}(t)+
\alpha(t)\frac{M_{n}^{\prime^2}(t)}{4}\\
-\alpha^{\prime}(t)M_{n}^{\prime}(t)-\alpha(t)\frac{M_{n}^{\prime\prime}(t)}{2})\geq0
\end{aligned}
\end{equation}
By simple algebra, we can show that,
\vspace{-.1in}
\begin{equation}\label{equ41}
\begin{aligned}
\alpha^{\prime}(t)=(\frac{t^2+1}{t})\alpha(t)-\frac{C_{v_j}}{\sqrt{2\pi}t}
\end{aligned}
\end{equation}
and
\vspace{-.1in}
\begin{equation}\label{equ42}
\begin{aligned}
\alpha^{\prime\prime}(t)=(t^2+3)\alpha(t)-\frac{C_{v_j}}{\sqrt{2\pi}}
\end{aligned}
\end{equation}
By substituting~(\ref{equ41}) and (\ref{equ42}) in~$g^{\prime\prime}_n(t)$, we have
\begin{equation}\label{equ43}
\begin{aligned}
g^{\prime\prime}_n(t)=\frac{\alpha(t)e^{-\frac{1}{2}M_{n}(t)}}{t}[t^3-t^2M_{n}^{\prime}(t)~~~~~~~~~~~~
\\+(3+\frac{M_{n}^{\prime^2}(t)}{4}-\frac{M_{n}^{\prime\prime}(t)}{2})t
-M_{n}^{\prime}(t)+\frac{C_{v_j}}{\sqrt{2\pi}\alpha(t)}(M_{n}^{\prime}(t)-t)]
\end{aligned}
\end{equation}
Now, we show~(\ref{equ40}) for all $t\geq0$.

First, since $nC_{v_j}^2\leq\rho_{n_j}\leq~n^2C_{v_j}^2$, we can show that,
\begin{equation}\label{equ44}
\begin{aligned}
\frac{M_{n}^{\prime^2}(t)}{4}-\frac{M_{n}^{\prime\prime}(t)}{2}\geq~t^2-1
\end{aligned}
\end{equation}
Then, from the following upper and lower bounds~\cite{abramowitz1972handbook}
 \begin{equation}\label{equ45}
\begin{aligned}
\frac{2}{t+\sqrt{t^2+4}}\leq~e^{\frac{t^2}{2}}\int_{t}^{\infty}e^\frac{-u^2}{2}du\leq~\frac{2}{t+\sqrt{t^2+\frac{8}{\pi}}}.
\end{aligned}
\end{equation}
we have,
\begin{equation}\label{equ46}
\begin{aligned}
\frac{C_{v_j}}{\sqrt{2\pi}}[1-\frac{2t}{t+\sqrt{t^2+\frac{8}{\pi}}}]\leq\alpha(t)\leq\frac{C_{v_j}}{\sqrt{2\pi}}[1-\frac{2t}{t+\sqrt{t^2+4}}]
\end{aligned}
\end{equation}
which indicates $\alpha(t)\geq0$ and we can show that
\begin{equation}\label{equ47}
\begin{aligned}
\frac{C_{v_j}}{\sqrt{2\pi}\alpha(t)}\geq(\frac{t+\sqrt{t^2+4}}{2})^2\geq~t^2+1
\end{aligned}
\end{equation}
From~(\ref{equ47}) and~(\ref{equ44}), the following inequality holds,
\begin{equation}\label{equ48}
\begin{aligned}
g^{\prime\prime}_n(t)\geq\frac{\alpha(t)e^{-\frac{1}{2}M_{n}(t)}}{t}[t^3-t^2M_{n}^{\prime}(t)~~~~~~~~~~~~
\\+(3+t^2-1)t
-M_{n}^{\prime}(t)+(t^2+1)(M_{n}^{\prime}(t)-t)]\\
=\alpha(t)e^{-\frac{1}{2}M_{n}(t)}(t^2+1)\geq0
\end{aligned}
\end{equation}
Therefore, for all $t\geq0$, i.e., $\mu_{g_{ij}}\geq1$, $g_n(\mu_{g_{ij}})$ and consequently $f(\lambda_{g_{ij}},\mu_{g_{ij}})$ is convex. The convexity of the following function:
\begin{equation}\label{equ49}
\begin{aligned}
f(\lambda_{b_{ij}},\mu_{b_{ij}})\triangleq\lambda_{b_{ij}}P_{L}(\lambda_{b_{ij}},\mu_{b_{ij}})-TH_j,~\forall i\in N,~\forall j\in J,
\end{aligned}
\end{equation}
can be similarly be proven and we conclude the convexity of inequality constraints~(\ref{equ23}), (\ref{equ24}).

Now, we prove the concavity of the objective function. Note that the nonnegative weighted sum of concave functions is concave~\cite{boyd2009convex}. Also, the functions $-\lambda_{g_{ij}}P_{L}(\lambda_{g_{ij}},\mu_{g_{ij}})$ and $-\lambda_{b_{ij}}P_{L}(\lambda_{b_{ij}},\mu_{b_{ij}})$ are concave. Therefore, by rewriting the objective functions based on
$-\lambda_{g_{ij}}P_{L}(\lambda_{g_{ij}},\mu_{g_{ij}})$ and $-\lambda_{b_{ij}}P_{L}(\lambda_{b_{ij}},\mu_{b_{ij}})$, we can show that if the data centers are profitable for each class of service, i.e.,
\begin{equation}\label{equ50}
\begin{aligned}
\negmedspace\delta_j+\gamma_j-\frac{P_{peak}-P_{idle}}{k_j}\max(C_{b_i},C_{g_i})\\
\geq~\delta_j-\frac{P_{peak}-P_{idle}}{k_j}\max(C_{b_i},C_{g_i})>0, \forall~i
\end{aligned}
\end{equation}
the objective function is concave and the proof is complete.
\end{IEEEproof}
\end{document}